\documentclass[twocolumn,showpacs,amssymb,eqsecnum]{revtex4}

\usepackage{dcolumn}
\usepackage{latexsym,epsf}

\begin{document}

\title{Vacuum solutions for scalar fields confined in cavities}

\author{David J. Toms}\email{d.j.toms@newcastle.ac.uk}
\affiliation{School of Mathematics, University of Newcastle Upon Tyne,\\
Newcastle Upon Tyne, United Kingdom NE1 7RU}

\date{\today}

\begin{abstract}
We look at vacuum solutions for fields confined in cavities where
the boundary conditions can rule out constant field
configurations, other than the zero field. If the zero field is
unstable, symmetry breaking can occur to a field configuration of
lower energy which is not constant. The stability of the zero
field is determined by the size of the length scales which
characterize the cavity and parameters that enter the scalar field
potential. There can be a critical length at which an instability
of the zero field sets in. In addition to looking at the
rectangular and spherical cavity in detail, we describe a general
method which can be used to find approximate analytical solutions
when the length scales of the cavity are close to the critical
value.
\end{abstract}

 \maketitle

\def\bea{\begin{eqnarray}}
\def\eea{\end{eqnarray}}
\def\be{\begin{equation}}
\def\ee{\end{equation}}
\def\nn{\nonumber\\}
\def\bp{\bar{\Psi}}
\def\b2{|\bp|^2}
\def\x{\mathbf{x}}
\def\bdry{\partial\Sigma}
\def\intS{\int_{\Sigma}\,d\sigma_x}
\def\sn{\mbox{\rm sn}}
\def\nn{\nonumber}
\def\Si{\mbox{\rm Si}}
\def\Ci{\mbox{\rm Ci}}

\section{Introduction}
\label{sec1}

The idea of spontaneous symmetry breaking plays a central role in
our current understanding of the standard model. (For an excellent
treatment by one of the major contributors see
Ref.~\cite{Weinberg} for example.) A toy model potential
demonstrating the most basic feature is the double-well potential
for a real scalar field considered by Goldstone~\cite{Goldstone}.
(See Eq.~(\ref{1.2}) below.) This potential has minimum away from
$\varphi=0$ at non-zero values of $\varphi=\pm a$, and it is easy
to show that these non-zero values represent possible stable
ground states for the theory.

For flat Minkowski spacetime, the homogeneity and isotropy of the
spacetime imply that the ground state must be constant.  However
it is easy to envisage a situation in which this is no longer the
case, even for homogeneous and isotropic spacetimes.  A simple
example was provided by Avis and Isham~\cite{AvisIsham} where the
spatial section of the spacetime was a torus.  In this case, since
a real scalar field is regarded as a cross-section of the
real-line bundle with the spacetime as the base space, there is
the possibility of a twisted field.  An equivalent, but simpler,
observation is that we could choose antiperiodic boundary
conditions for the field on the torus just as well as periodic
boundary conditions. Antiperiodic boundary conditions mean that
the only constant field solution consistent with the boundary
conditions is the zero field. For the double-well potential, this
raises the interesting question of what the ground state is if
cannot correspond to a minimum of the potential.  What Avis and
Isham~\cite{AvisIsham} showed was that if the torus was below a
critical size set by the parameters of the double-well potential,
$\varphi=0$ was the stable ground state; however, if the torus was
larger than the critical size, $\varphi=0$ was not stable, and the
stable ground state was a spatially dependent solution.  Thus the
boundary conditions can prohibit constant, non-zero field
solutions, and make the determination of the ground state more
complicated.

The main purpose of the present paper is to show that the
situation for scalar fields confined in cavities is similar to
that studied by Avis and Isham~\cite{AvisIsham} for twisted scalar
fields. In Sec.~\ref{sec2} we will present a fairly general
analysis showing how to characterise the critical size of cavities
corresponding to the stability (or instability) of the zero field
as the ground state. In situations where $\varphi=0$ is unstable,
an exact determination of the ground state is not possible in
general. However, when the length scales of the cavity are close
to the critical values at which an instability of $\varphi=0$ sets
in, it is possible to find approximate analytical solutions; this
is described in Sec.~\ref{sec3}. One case where an exact
analytical solution is possible occurs for a rectangular cavity
where the field vanishes on only two opposite pairs of the cavity
walls. This case is studied in Sec.~\ref{exact1D}.  In
Sec.~\ref{approxcavity} we use the approximate method of
Sec.~\ref{sec3} to obtain a solution which is shown to agree with
an expansion of the exact solution we found in Sec.~\ref{exact1D}.
In Sec.~\ref{Dirichlet3D} we study the case where the field is
totally confined by the cavity, vanishing on all of the box walls.
The case of a spherical cavity, with the field vanishing on the
surface of a spherical shell, is examined in Sec.~\ref{sec5}.  No
exact solution could be found, but we obtain a numerical solution,
as well as apply the approximate method of Sec.~\ref{sec3}.  The
final section contains a brief discussion of the results.

\section{General framework}
\label{sec2}

We will consider a general region of $D$-dimensional Euclidean
space $\Sigma$ with boundary $\bdry$. (The basic formalism we will
describe does not require the space to be flat, but we will not
consider curved space in this paper.) The action functional for a
real scalar field $\varphi$ will be chosen to be
\begin{equation}
S\lbrack\varphi\rbrack=\int\,dt\intS\left\lbrace
\frac{1}{2}\partial^\mu\varphi\partial_\mu\varphi-U(\varphi)\right\rbrace\;,
\label{1.1}
\end{equation}
where $U(\varphi)$ is some potential, and $d\sigma_x$ denotes the
invariant volume element for $\Sigma$ which of course depends on
the choice of coordinates. We will concentrate on the case of the
double-well potential
\begin{equation}
U(\varphi)=\frac{\lambda}{4!}(\varphi^2-a^2)^2\label{1.2}
\end{equation}
in this paper, although the analysis is easily extended to other
potentials in a straightforward manner. The field equation which
follows from Eq.~(\ref{1.1}) is
\begin{equation}
\Box\varphi+U'(\varphi)=0\;.\label{1.3}
\end{equation}
The aim is to solve this for $\varphi$ in the region $\Sigma$,
subject to some boundary conditions imposed on $\varphi$ on the
boundary, $\bdry$.

The energy functional (which is just the spatial integral of the
Hamiltonian density) is
\begin{equation}
E\lbrack\varphi\rbrack=\intS\left\lbrace
\frac{1}{2}\dot{\varphi}^2+\frac{1}{2}|\nabla\varphi|^2
+U(\varphi)\right\rbrace\;. \label{1.4}
\end{equation}
Because the spacetime is static, if we concentrate on the lowest
energy solution, or ground state, we would expect it to be static
as well, resulting in no contribution to the kinetic energy term
in Eq.~(\ref{1.4}). For static solutions Eq.~(\ref{1.3}) becomes
\begin{equation}
-\nabla^2\varphi+U'(\varphi)=0\;.\label{1.5}
\end{equation}
If we assume that $\varphi$ has no time dependence, then
Eq.~(\ref{1.4}) reads
\begin{equation}
E\lbrack\varphi\rbrack=\intS\left\lbrace
\frac{1}{2}\varphi(-\nabla^2)\varphi+U(\varphi)\right\rbrace\;,
\label{1.6}
\end{equation}
after an integration by parts, provided that the boundary
conditions are such that $\varphi\nabla_n\varphi$ vanishes on
$\bdry$, with $\nabla_n$ the outwards directed normal derivative.
({\em ie}\/ $\nabla_n=\hat{n}\cdot{\nabla}$ with $\hat{n}$ a unit
normal vector on $\bdry$ which is directed out of the boundary.)
We will assume that this is the case here.

For the potential Eq.~(\ref{1.2}), it is obvious that $\varphi=0$
and $\varphi=\pm a$ are solutions to Eq.~(\ref{1.5}). There are
two key issues that arise at this stage. The first is whether or
not the boundary conditions on $\bdry$ are satisfied; if not, then
the solution must be rejected. The second is, even if the boundary
conditions are obeyed, the solution must be stable to
perturbations. Assuming that we have a solution $\varphi$ to
Eq.~(\ref{1.5}) which satisfies the boundary conditions on
$\bdry$, we can look at the energy of a perturbed solution
$\varphi+\psi$ where $\psi$ is treated as small. From
Eq.~(\ref{1.6}) we find
\begin{equation}
E\lbrack\varphi+\psi\rbrack-E\lbrack\varphi\rbrack=\frac{1}{2}
\int d\sigma_x\psi\left(-\nabla^2+U''(\varphi)\right)\psi
\label{1.7}
\end{equation}
if we work to lowest order in $\psi$. By considering the
eigenvalue problem
\begin{equation}
\left(-\nabla^2+U''(\varphi)\right)\psi_n=\lambda_n\psi_n
\;,\label{1.8}
\end{equation}
it can be seen that if any of the eigenvalues $\lambda_n$ are
negative, then
$E\lbrack\varphi+\psi_n\rbrack<E\lbrack\varphi\rbrack$ and we
would conclude that the solution $\varphi$ is unstable to small
perturbations since a solution of lower energy exists. If all of
the eigenvalues are positive, then $\varphi$ is stable to small
perturbations (locally stable).

Suppose that we consider the solution $\varphi=0$ to
Eq.~(\ref{1.5}) with the potential Eq.~(\ref{1.2}). The eigenvalue
equation Eq.~(\ref{1.8}) becomes
\begin{equation}
\left(-\nabla^2-\frac{\lambda}{6}a^2\right)\psi_n=\lambda_n\psi_n
\;.\label{1.9}
\end{equation}
If we consider the lowest eigenvalue $\lambda_0$, whose sign
determines the stability of $\varphi=0$, this sign requires
knowing the smallest eigenvalue of the Laplacian $-\nabla^2$. Let
\begin{equation}
-\nabla^2\psi_0=\ell_0^2\psi_0\;,\label{1.10}
\end{equation}
where $\ell_0^2\ge0$ and is real. We then have
\begin{equation}
\lambda_0=\ell_0^2-\frac{\lambda}{6}a^2\;.\label{1.11}
\end{equation}
If the boundary conditions on $\varphi$ (and hence $\psi_n$) are
such that $\ell_0^2=0$, then we can conclude that $\lambda_0<0$
and so $\varphi=0$ is unstable. This is the case where the
boundary conditions allow constant values of the field, and is the
case for flat Euclidean space. However, if the boundary conditions
prohibit constant values of the field, we will have $\ell_0^2>0$,
and the stability or instability of $\varphi=0$ is determined by
the magnitude of $\ell_0^2$ in relation to $\frac{\lambda}{6}a^2$.
Typically $\ell_0^2$ will depend on the length scales describing
$\Sigma$ and its boundary, and so we will find critical values of
these lengths at which $\varphi=0$ becomes unstable. The ground
state will not be constant in this case. The first demonstration
of this was given in Ref.~\cite{AvisIsham} for the case where
$\Sigma$ was a circle, or a torus, with the field $\varphi$
satisfying antiperiodic boundary conditions. We will see examples
of this for fields in cavities later in this paper.

The other constant solutions to Eq.~(\ref{1.5}) are $\varphi=\pm
a$. The eigenvalue equation Eq.~(\ref{1.8}) reads
\begin{equation}
\left(-\nabla^2+\frac{\lambda}{3}a^2\right)\psi_n=\lambda_n\psi_n
\label{1.12}
\end{equation}
for these solutions. In place of Eq.~(\ref{1.11}) we have
\begin{equation}
\lambda_0=\ell_0^2+\frac{\lambda}{3}a^2\label{1.13}
\end{equation}
with $\ell_0^2$ defined as in Eq.~(\ref{1.10}). Provided that the
boundary conditions allow constant values of the field, if
$\varphi=0$ is unstable we see that $\lambda_0>0$ implying
stability of the solutions $\varphi=\pm a$. This is the usual
situation in Euclidean space~\cite{Goldstone}. However if the
boundary conditions rule out $\varphi=\pm a$ as possible
solutions, then the ground state when $\varphi=0$ is unstable
necessarily involves non-constant fields.

A simple example where constant field values (other than zero) are
not allowed occurs for a field confined to a cavity $\Sigma$ with
Dirichlet boundary conditions imposed on the boundary $\bdry$ of
the cavity. ({\em ie}\/ $\varphi=0$ on $\bdry$.) In this case, if
$\varphi=0$ is not stable we are faced with the question of what
the ground state is. In Sec.~\ref{sec4} we will look at the
simplest case of a cavity which is rectangular. In Sec.~\ref{sec5}
we will study what happens for a spherical cavity. The case of the
rectangular box allows for an exact solution for the ground state
for some boundary conditions; whereas for the spherical cavity we
were not able to solve for the ground state except numerically.
Before proceeding to these examples, we will present a method for
obtaining approximate analytical solutions for the ground state.

\section{Approximate analytical results}
\label{sec3}

In Ref.~\cite{LauraToms} a systematic method was described for
obtaining approximate solutions for the ground state in cases
where it was not possible to solve Eq.~(\ref{1.5}) exactly (except
numerically). In this section we will present a variation on this
method applicable to general cavities. We saw that the stability
of $\varphi=0$ relied on the magnitude of $\ell_0^2$ which in turn
depends on the length scales of the cavity. For simplicity assume
that there is only one length scale $L$ in the problem. Let $L_c$
be the critical value of this length defined by the condition
\begin{equation}
\ell_0^2=\frac{\lambda}{6}a^2\;.\label{2.1}
\end{equation}
On dimensional grounds, we must have $\ell_0^2\propto L^{-2}$, so
for $L<L_c$ we expect $\lambda_0>0$ and hence for $\varphi=0$ to
be locally stable. When $L>L_c$, we have $\lambda_0<0$, and so
$\varphi=0$ becomes unstable. In this case the ground state is
given by a solution other than $\varphi=0$.

Suppose that we take the length scale $L$ to be slightly greater
than the critical value~:
\begin{equation}
L=(1+\epsilon)L_c\;,\label{2.2}
\end{equation}
with $\epsilon>0$ treated as small. We can scale the coordinates
with appropriate factors of $L$ to enable us to use dimensionless
coordinates. The Laplacian $-\nabla^2$ can then be expanded in
powers of $\epsilon$ using Eq.~(\ref{2.2}) to give
\begin{equation}
-\nabla^2=-\nabla_c^2-\epsilon\nabla_1-\epsilon^2\nabla_2-\cdots
\;,\label{2.3}
\end{equation}
where $\nabla_c^2$ denotes that the Laplacian is evaluated with
$L=L_c$, and $\nabla_1,\nabla_2,\ldots$ are differential operators
whose form is determined by the Laplacian and the specific cavity.
(We will apply this to the rectangular box and the spherical
cavity later to show how this works in practice.)

An argument presented in Ref.~\cite{LauraToms}, which is
substantiated by application to specific examples, may be repeated
to show that the solution to Eq.~(\ref{1.5}) can be written as
\begin{equation}
\varphi(x)=\epsilon^{1/2}\left\lbrack\varphi_0(x)
+\epsilon\varphi_1(x)+\epsilon^2\varphi_2(x)
+\cdots\right\rbrack\;,\label{2.4}
\end{equation}
where $\varphi_0(x),\varphi_1(x),\ldots$ are independent of
$\epsilon$. Using Eqs.~(\ref{2.3}) and (\ref{2.4}) in
Eq.~(\ref{1.5}), with the potential given by Eq.~(\ref{1.2}), and
then equating equal powers of $\epsilon$ to zero, we obtain a set
of coupled differential equations, the first two of which are
\begin{eqnarray}
-\nabla_c^2\varphi_0-\frac{\lambda}{6}a^2\varphi_0&=&0\;,\label{2.5}\\
-\nabla_c^2\varphi_1-\frac{\lambda}{6}a^2\varphi_1&=&\nabla_1\varphi_0
-\frac{\lambda}{6}\varphi_0^3\;.\label{2.6}
\end{eqnarray}
(Higher order equations may be obtained in a straightforward
manner if needed.) The set of differential equations obtained may
be solved iteratively beginning with Eq.~(\ref{2.5}). By comparing
Eq.~(\ref{2.5}) with Eq.~(\ref{1.10}) where $\ell_0^2$ is defined
by the critical length in Eq.~(\ref{2.1}), it is observed that
$\varphi_0$ is the eigenfunction of the Laplacian whose eigenvalue
is the smallest (with $L$ set equal to $L_c$).

Because Eq.~(\ref{2.5}) is a homogeneous equation, the overall
scale of the solution is not determined. However Eq.~(\ref{2.6}),
as well as the higher order equations not written down explicitly,
are inhomogeneous. The scale of $\varphi_0$ can be fixed by
requiring Eq.~(\ref{2.4}) to minimize the energy to lowest order
in $\epsilon$. From Eq.~(\ref{1.6}), using Eq.~(\ref{2.4}) along
with Eqs.~(\ref{2.5}) and (\ref{2.6}), it follows that
\begin{equation}
E=\intS\left\lbrace
\frac{\lambda}{24}a^4-\frac{1}{2}\epsilon^2\varphi_0\nabla_1\varphi_0
+\frac{\lambda}{24}\epsilon^2\varphi_0^4
\right\rbrace\;,\label{2.7}
\end{equation}
where we have kept only the lowest order terms in $\epsilon$. If
we let $\tilde{\varphi}_0$ be any solution to Eq.~(\ref{2.5})
which satisfies the correct boundary conditions, and write
\begin{equation}
\varphi_0=A\tilde{\varphi}_0\label{2.8}
\end{equation}
for some constant $A$, we obtain
\begin{equation}
E=C_0-C_1A^2+C_2A^4\label{2.9}
\end{equation}
where
\begin{equation}
C_0=\intS\frac{\lambda}{24}a^4\label{2.10}
\end{equation}
is independent of $A$ and $\epsilon$, and
\begin{eqnarray}
C_1&=&\frac{1}{2}\epsilon^2\intS\tilde{\varphi}_0\nabla_1\tilde{\varphi}_0
\;,\label{2.11}\\
C_2&=&\frac{\lambda}{24}\epsilon^2\intS\tilde{\varphi}_0^4\;,\label{2.12}
\end{eqnarray}
are independent of $A$. For $C_1>0$, we have
\begin{equation}
A=\pm\left(\frac{C_1}{2C_2}\right)^{1/2}\label{2.13}
\end{equation}
as the value of $A$ which makes the energy a local minimum. This
sets the scale of the solution for $\varphi_0$~:
\begin{equation}
\varphi_0=\pm\left(\frac{C_1}{2C_2}\right)^{1/2}
\tilde{\varphi}_0\label{2.14}
\end{equation}
with $\tilde{\varphi}_0$ any solution to Eq.~(\ref{2.5}).

To proceed to the next order in $\epsilon$ we must solve
Eq.~(\ref{2.6}) with $\varphi_0$ given by Eq.~(\ref{2.14}). The
general solution may be expressed as
\begin{equation}
\varphi_1=\varphi_{1h}+\varphi_{1p}\label{2.15}
\end{equation}
where $\varphi_{1h}$ is a solution of the homogeneous equation
Eq.~(\ref{2.5}) and $\varphi_{1p}$ is any particular solution to
Eq.~(\ref{2.6}). The overall scale of $\varphi_{1p}$ is fixed
since it satisfies an inhomogeneous equation; however, the scale
of $\varphi_{1h}$ is not determined. We can choose $\varphi_{1h}$
to be proportional to $\tilde{\varphi}_0$ and fix the constant of
proportionality by requiring that the energy be minimized to the
next order in $\epsilon$. We therefore take
\begin{eqnarray}
\varphi(x)&=&\epsilon^{1/2}\Big\lbrack
\left(\frac{C_1}{2C_2}\right)^{1/2}\tilde{\varphi}_0
+\epsilon(A\tilde{\varphi}_0+\varphi_{1p})\nonumber\\&&\quad+
\epsilon^2\varphi_2+\cdots\Big\rbrack\label{2.16}
\end{eqnarray}
in Eq.~(\ref{1.6}) and minimize the resulting expression for
$E(A)$. Because we are only interested in the value of $A$ which
makes $E(A)$ a minimum, it is simpler to evaluate
\begin{equation}
\frac{\partial}{\partial
A}E(A)=\epsilon^{3/2}\intS\tilde{\varphi}_0 (-\nabla^2\varphi+
\frac{\lambda}{6}\varphi^3-\frac{\lambda}{6}a^2\varphi)
\label{2.17}
\end{equation}
to lowest order in $\epsilon$. Because we have only solved the
equation of motion up to and including terms of order
$\epsilon^{3/2}$ in obtaining Eqs.~(\ref{2.5}) and (\ref{2.6}),
the next term in the integrand of Eq.~(\ref{2.17}) will be of
order $\epsilon^{5/2}$. A short calculation shows that
\begin{equation}
\frac{\partial}{\partial A}E(A)=\epsilon^{4}(D_0+D_1A)\label{2.18}
\end{equation}
where
\begin{eqnarray}
D_0&=&\intS\Big\lbrack-\tilde{\varphi}_0\nabla_1\varphi_{1p}-
\left(\frac{C_1}{2C_2}\right)^{1/2}\tilde{\varphi}_0
\nabla_2\tilde{\varphi}_0\nonumber\\
&&\quad\quad+\frac{\lambda}{2}\left(\frac{C_1}{2C_2}\right)
\tilde{\varphi}_0^3\varphi_{1p}\Big\rbrack\;,\label{2.19}\\
D_1&=&\intS\Big\lbrack-\tilde{\varphi}_0\nabla_1\tilde{\varphi}_0
+\frac{\lambda}{2}\left(\frac{C_1}{2C_2}\right)
\tilde{\varphi}_0^4\Big\rbrack\;.\label{2.20}
\end{eqnarray}
We conclude that if $D_1>0$, then
\begin{equation}
A=-\frac{D_0}{D_1}\label{2.21}
\end{equation}
gives the value which minimizes $E(A)$ to lowest order in
$\epsilon$. From Eq.~(\ref{2.16}), in summary, we therefore have
the approximate solution given to order $\epsilon^{3/2}$ by
\begin{equation}
\varphi(x)=\epsilon^{1/2}\Big\lbrack
\left(\frac{C_1}{2C_2}\right)^{1/2}\tilde{\varphi}_0
+\epsilon(-\frac{D_0}{D_1}\tilde{\varphi}_0+\varphi_{1p})
\Big\rbrack\;,\label{2.22}
\end{equation}
where $\tilde{\varphi}_0$ and $\varphi_{1p}$ are any solutions to
Eqs.~(\ref{2.5}) and (\ref{2.6}) respectively. $C_1$ and $C_2$ are
defined by Eqs.~(\ref{2.11}) and (\ref{2.12}). $D_0$ and $D_1$ are
defined by Eqs.~(\ref{2.19}) and (\ref{2.20}).

It should be clear from the analysis that we have presented how
the method extends to any order in $\epsilon$. First solve the
equation of motion Eq.~(\ref{1.3}) to order $\epsilon^{n+1/2}$
using Eq.~(\ref{2.4}) extended to order $\epsilon^{n+1/2}$ and
Eq.~(\ref{2.3}) to order $\epsilon^{n}$. ($n=0,1,2,\ldots$ here.)
Evaluate $\frac{\partial}{\partial A}E(A)$ to order
$\epsilon^{2n+2}$ using the solutions found, and then solve for
$A$ as we have illustrated for the cases $n=0,1$.

\section{The rectangular cavity}
\label{sec4}
\subsection{Exact result}
\label{exact1D}

The simplest case of a cavity where an exact solution can be found
occurs for a field confined inside of a rectangular box. The first
case we will look at is for a field which satisfies a Dirichlet
boundary condition on opposite sides of one pair of box walls. We
will choose this to be the $y$ direction and take $-L/2\le y\le
L/2$ with $\varphi=0$ when $y=\pm L/2$. In the $x$ and $z$
directions we will choose either periodic boundary conditions, or
else Neumann boundary conditions with $\frac{\partial}{\partial
x}\varphi$ and $\frac{\partial}{\partial z}\varphi$ vanishing at
$x=\pm L_x/2$ and $z=\pm L_z/2$ respectively. With either of these
two choices, the ground state will not depend on the $x$ or $z$
coordinates, and the problem reduces to one that is
one-dimensional. One physical application of this is to the case
of two parallel plates, as in the Casimir effect, where the plate
separation is much less than their linear extent. ({\em ie}\/ Keep
$L$ finite, and let $L_x,L_z\rightarrow\infty$.) In this case,
with $L_x,L_z\rightarrow\infty$, the choice of boundary conditions
in the $x$ and $z$ directions would not be expected to be
important. Later in this section we will discuss what happens in
the case where Dirichlet boundary conditions are imposed in all
three spatial directions.

If we assume $\varphi=\varphi(y)$ only, then Eq.~(\ref{1.5})
becomes
\begin{equation}
-\frac{d^2\varphi}{dy^2}+\frac{\lambda}{6}
\varphi(\varphi^2-a^2)=0\;.\label{3.1}
\end{equation}
This equation admits a first integral,
\begin{equation}
-\frac{1}{2}\left(\frac{d\varphi}{dy}\right)^2
+\frac{\lambda}{24}(\varphi^2-a^2)^2=C \;,\label{3.2}
\end{equation}
with $C$ a constant. Note that $\varphi=\pm a$ solves
Eq.~(\ref{3.1}), but does not satisfy the requirement that the
field vanish at $\varphi=\pm L/2$, so is not allowed. $\varphi=0$
satisfies Eq.~(\ref{3.1}) as well as the boundary conditions, so
is a valid solution. To see if $\varphi=0$ is locally stable we
look for the solution to Eq.~(\ref{1.10}) of lowest eigenvalue.
Because $\psi_0(y=\pm L/2)=0$, we have
$\psi_0\propto\sin\left(\frac{\pi}{L}(y+L/2)\right)$ and
$\ell_0^2=\pi^2/L^2$. From Eq.~(\ref{1.11}) we see that
$\varphi=0$ is stable if $L<L_c$ where
\begin{equation}
L_c=\frac{\pi}{a}\left(\frac{6}{\lambda}\right)^{1/2}
\;.\label{3.3}
\end{equation}
If $L>L_c$, then $\varphi=0$ is unstable, and because the boundary
conditions prohibit constant values of $\varphi$, the ground state
must be spatially dependent. We will now find this solution.

For $\varphi(y)$ to be continuous on the interval $\lbrack
-L/2,L/2\rbrack$ with $\varphi(\pm L/2)=0$, it must have a
stationary value somewhere. This allows us to conclude that $C$
defined in Eq.~(\ref{3.2}) must be non-negative, and also that
$C\le\frac{\lambda}{24}a^4$. We will define
\begin{equation}
C=\frac{\lambda}{24}a^4w^2\;,\label{3.4}
\end{equation}
with $w$ real and satisfying $0\le w\le1$. The cases $w=0$ and
$w=1$ require special treatment. For $w=1$ it is easy to show that
the only solution to Eq.~(\ref{3.2}) which satisfies the boundary
conditions that $\varphi=0$ at $y=\pm L/2$ is $\varphi=0$ for all
$y$. We already know that this solution is unstable for $L>L_c$ so
we will concentrate on $w<1$. The case $w=0$ is simple to solve,
and leads to the usual kink solution. (See
Ref.~\cite{Coleman,Jackiw} for example.) This does not satisfy our
boundary conditions. Therefore we will restrict $0<w<1$. It is
worth remarking on the difference between the confined case and
the field in an unbounded region. In the latter case, the
requirement that the field configuration leads to a finite energy
requires the field to asymptotically approach a zero of the
potential~\cite{Coleman,Jackiw}. For a field confined by a cavity,
any well-behaved function will have a finite energy as a trivial
consequence of the finite volume of the region; thus, the finite
energy requirement does not give anything useful in our case.

If we restrict $0<w<1$, then the solution to Eq.~(\ref{3.2}) can
be expressed in terms of Jacobi elliptic functions~\cite{WW}. We
find
\begin{equation}
\varphi(y)=\pm a(1-w)^{1/2}\sn\left(
\beta\left(y+\frac{L}{2}\right),\left(\frac{1-w}{1+w}\right)^{1/2}
\right)\;,\label{3.5}
\end{equation}
where
\begin{equation}
\beta=\left\lbrack\frac{\lambda}{12}
(1+w)a^2\right\rbrack^{1/2}\;,\label{3.6}
\end{equation}
as the solution to Eq.~(\ref{3.2}) which vanishes at $y=-L/2$. If
we impose the other boundary condition at $y=L/2$, we find the
condition ($w\ne1$)
\begin{equation}
\sn\left(\beta L,\left(\frac{1-w}{1+w}\right)^{1/2}\right)=0
\;.\label{3.7}
\end{equation}
Making use of the property~\cite{WW} that $\sn(u,k)$ has zeros for
$u=2nK(k)$ for $n=0,\pm1,\pm2,\ldots$ where $K(k)$ is the complete
elliptic integral of the first kind,
\begin{equation}
K(k)=\int_{0}^{\pi/2}d\phi(1-k^2\sin^2\phi)^{-1/2} \;,\label{3.8}
\end{equation}
we conclude that
\begin{equation}
\beta L=2nK\left(\left(\frac{1-w}{1+w}\right)^{1/2}\right)
\;,\label{3.9}
\end{equation}
where $n=1,2,\ldots$. We can restrict $n$ to non-negative values
since $\sn(u,k)$ is an odd function of $u$, and a sign change of
$u$ just changes the overall sign of $\varphi$ which is arbitrary.
We rule out $n=0$ because this corresponds to $\varphi$
identically zero in Eq.~(\ref{3.5}), and we already know that this
is unstable for $L>L_c$. The interpretation of $n$ is that $(n-1)$
gives the number of zeros of $\varphi$ in the open interval
$(-L/2,L/2)$.

The next question concerns the solution for the arbitrary
integration constant $w$ in Eq.~(\ref{3.9}). We need to solve
\begin{equation}
\gamma_n=(1+w)^{-1/2}K\left(\left(\frac{1-w}{1+w}\right)^{1/2}\right)
\;,\label{3.10}
\end{equation}
where
\begin{equation}
\gamma_n=\frac{La}{2n}\left(\frac{\lambda}{12}\right)^{1/2}
\;,\label{3.11}
\end{equation}
for $w$, with $n=1,2,\ldots$. The properties of the complete
elliptic integral of the first kind may be used to show that the
right hand side of Eq.~(\ref{3.10}) is a monotonically decreasing
function of $w$ on the interval $0\le w\le1$, approaching infinity
as $w\rightarrow0$ and the value $\pi\sqrt{2}/4$ as
$w\rightarrow1$. This means that a solution to Eq.~(\ref{3.10})
only exists if
\begin{equation}
\gamma_n>\frac{\pi\sqrt{2}}{4}\;.\label{3.12}
\end{equation}
Making use of Eq.~(\ref{3.11}) and the definition of the critical
length in Eq.~(\ref{3.3}) shows that we will always have a
solution for $w$ in Eq.~(\ref{3.10}) if
\begin{equation}
L>nL_c\;.\label{3.13}
\end{equation}
Apart from the different boundary conditions, the situation is
very similar to the twisted field case on the circle examined by
Avis and Isham~\cite{AvisIsham}. In fact, we can make use of their
clever proof that for $n\ge2$ the solutions for $\varphi(y)$ are
all unstable. This is not unexpected, since the energy of
solutions with $n\ge2$ are all greater than that for the $n=1$
solution. We will therefore concentrate on the case of $n=1$.

If we use the definition of $L_c$ given in Eq.~(\ref{3.3}) in
Eq.~(\ref{3.10}), when $n=1$ we find
\begin{equation}
\frac{\pi L}{2\sqrt{2}L_c}\sqrt{1+w}=
K\left(\sqrt{\frac{1-w}{1+w}}\right) \;.\label{3.14}
\end{equation}
showing that the solution for $w$ depends only on the
dimensionless ratio of $L/L_c$. By expanding both sides of
Eq.~(\ref{3.14}) in powers of $w$, it is easy to show that for
large values of $L/L_c$ a good approximation for the solution is
given by
\begin{equation}
w\simeq 8\exp\left(-\frac{\pi L}{\sqrt{2}L_c}\right)
\;.\label{3.15}
\end{equation}
Even for relatively small values of $L/L_c$ this turns out to be
quite accurate. (For example, when $L/L_c=3$, we find an agreement
to six decimal places between this approximation and the result of
solving Eq.~(\ref{3.14}) numerically.)  The solution to
Eq.~(\ref{3.14}) is shown in Fig.~1 for a range of $L/L_c$.
\begin{figure}[htb]
\begin{center}
\leavevmode \epsfxsize=80mm \epsffile{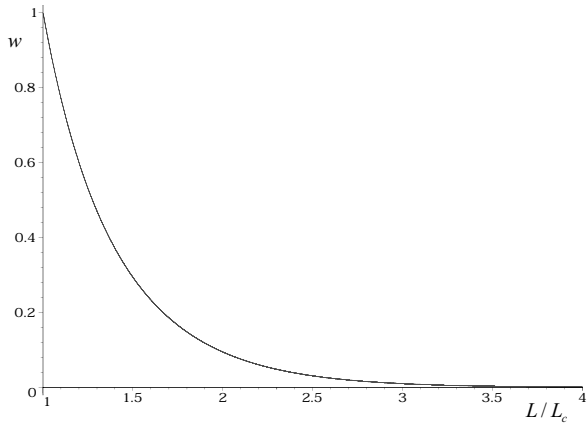}
\end{center}
\caption{The solution for $w$ obtained by solving Eq.~(\ref{3.14})
as a function of $L/L_c$.}
\end{figure}
It is clear from this figure that as the value of $L/L_c$ is
increased, $w$ tends towards 0, as predicted from the
approximation Eq.~(\ref{3.15}). Since $w=0$ corresponds to the
constant solution $\varphi=a$, it would be expected that as we
increase the value of the ratio $L/L_c$, the solution for
$\varphi$ will try to be as close as it can to the constant value
of $a$. This is in fact what happens as we show in Fig.~2.
\begin{figure}[htb]
\begin{center}
\leavevmode \epsfxsize=80mm \epsffile{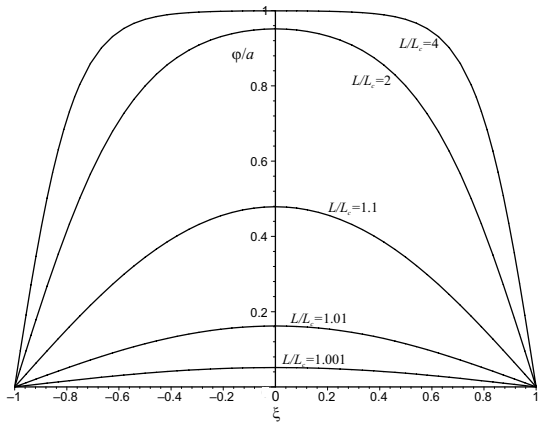}
\end{center}
\caption{The solution for $\varphi(\xi)/a$ is shown for various
values of $L/L_c$ as a function of $\xi=2y/L$.}
\end{figure}
As $L/L_c$ gets larger the solution tends towards the step
function, which is as close as the boundary conditions allow it to
get to the standard solution $\varphi=a$.

\subsection{Approximate result}
\label{approxcavity}

As a test of the approximation method described in
Sec.~\ref{sec3}, we will use it to compare with the exact solution
we have just found. Take $L$ to be close to the critical length
$L_c$ as in Eq.~(\ref{2.2}) with $L_c$ defined by Eq.~(\ref{3.3})
in the present case. As explained in Sec.~\ref{sec3}, it is
advantageous to adopt dimensionless coordinates, so we will define
\begin{equation}
y=\frac{L}{2}\xi\;,\label{4.2.1}
\end{equation}
with $-1\le\xi\le1$. Since
$\displaystyle{\nabla^2=\frac{d^2}{dy^2}}$, by using
Eq.~(\ref{4.2.1}) and the expansion of $L$ about $L_c$ as in
Eq.~(\ref{2.2}), it is easy to read off $\nabla_1$ and $\nabla_2$
defined in Eq.~(\ref{2.3}) to be
\begin{eqnarray}
\nabla_1&=&-\frac{8}{L_c^2}\,\frac{d^2}{d\xi^2}\;,\label{4.2.2a}\\
\nabla_2&=&\frac{12}{L_c^2}\,\frac{d^2}{d\xi^2}\;.\label{4.2.2b}\\
\end{eqnarray}
We also have
\begin{equation}
\nabla_c=\frac{4}{L_c^2}\,\frac{d^2}{d\xi^2}\;.\label{4.2.2c}
\end{equation}

The next step is to solve Eq.~(\ref{2.5}) for any solution
$\tilde{\varphi}_0$. Using Eq.~(\ref{4.2.2c}) and the definition
of $L_c$ in Eq.~(\ref{3.3}), it is easy to see that
\begin{equation}
\tilde{\varphi}_0(\xi)=\cos\left( \frac{\pi}{2}\xi
\right)\label{4.2.3}
\end{equation}
is a solution which satisfies the proper boundary conditions. A
straightforward calculation of $C_1$ and $C_2$ defined by
Eqs.~(\ref{2.11}) and (\ref{2.12}) leads to the conclusion that
\begin{equation}
\frac{C_1}{2C_2}=\frac{8}{3}a^2\;.\label{4.2.4}
\end{equation}
We therefore have as our leading order approximation to the ground
state
\begin{equation}
\varphi(\xi)\simeq\left(\frac{8}{3}\,\epsilon\right)^{1/2}
a\cos\left( \frac{\pi}{2}\xi \right)\;.\label{4.2.5}
\end{equation}

Before comparing this with the expansion of the exact result, we
will first evaluate the next order correction to
Eq.~(\ref{4.2.5}). This entails initially solving  Eq.~(\ref{2.6})
for any solution $\varphi_{1p}$. With
\begin{equation}
\varphi_0(\xi)=\left(\frac{8}{3}\right)^{1/2}a\cos\left(
\frac{\pi}{2}\xi \right)\;,\label{4.2.6}
\end{equation}
and $\nabla_1$ given by Eq.~(\ref{4.2.2a}), it can be shown that a
particular solution to Eq.~(\ref{2.6}) is given by
\begin{equation}
\varphi_{1p}(\xi)=-\frac{\sqrt{6}}{18}\,a\cos\left(
\frac{3\pi}{2}\xi \right)\;.\label{4.2.7}
\end{equation}
The constants $D_0$ and $D_1$ defined in Eqs.~(\ref{2.19}) and
(\ref{2.20}) may now be evaluated with the result
\begin{equation}
\frac{D_0}{D_1}=\frac{17\sqrt{6}}{36}\,a\;.\label{4.2.8}
\end{equation}
Combining all of these results, and using Eq.~(\ref{2.22}), we
obtain the approximate ground state solution to be
\begin{eqnarray}
\hspace{-12pt}\varphi(\xi)&\simeq&\epsilon^{1/2}\left\lbrace
\left(\frac{8}{3}\right)^{1/2}a\cos\left( \frac{\pi}{2}\xi
\right)\right.\nn\\
&&\hspace{-36pt}\left.-\epsilon\left\lbrack\frac{17\sqrt{6}}{36}\,a\cos\left(
\frac{\pi}{2}\xi \right) +\frac{\sqrt{6}}{18}\,a\cos\left(
\frac{3\pi}{2}\xi \right)\right\rbrack\right\rbrace,\label{4.2.9}
\end{eqnarray}
up to, and including, terms of order $\epsilon^{3/2}$. (We drop
the $\pm$ here.)

The exact ground state solution was found to be given by
Eqs.~(\ref{3.5},\ref{3.6},\ref{3.9}) with $n=1$. Using the
dimensionless coordinate $\xi$ defined in Eq.~(\ref{4.2.1}) the
exact solution reads
\begin{equation}
\varphi(\xi)=a(1-w)^{1/2}\sn\Big((1+\xi)K(k_w),k_w\Big)\;,\label{4.2.10}
\end{equation}
where
\begin{equation}
k_w=\left(\frac{1-w}{1+w}\right)^{1/2}\;.\label{4.2.11}
\end{equation}
We will now expand this result consistently to order
$\epsilon^{3/2}$ with $L=(1+\epsilon)L_c$.

For $L$ close to $L_c$, $w$ will be close to 1; thus we will let
\begin{equation}
w=1-\eta\;,\label{4.2.12}
\end{equation}
with $\eta<<1$. The first task is to determine $\eta$ in terms of
$\epsilon$. From Eqs.~(\ref{3.10}) and (\ref{3.11}), with $n=1$,
it can be shown that
\begin{equation}
1+\epsilon=\frac{4}{\pi\sqrt{2}}(1+w)^{-1/2}K(k_w)\;.\label{4.2.13}
\end{equation}
(The result has been simplified using Eq.~(\ref{3.3}).) Using
Eq.~(\ref{4.2.12}) in Eq.~(\ref{4.2.11}), and expanding the
complete elliptic integral of the first kind in powers of $\eta$
results in
\begin{equation}
1+\epsilon\simeq1+\frac{3}{8}\eta+\frac{57}{256}\eta^2
+\cdots\;.\label{4.2.14}
\end{equation}
This expansion may be inverted to find
\begin{equation}
\eta\simeq\frac{8}{3}\epsilon-\frac{38}{9}\epsilon^2
+\cdots\;,\label{4.2.15}
\end{equation}
which when used in Eq.~(\ref{4.2.12}), gives us an expansion of
the parameter $w$ in terms of $\epsilon$.

The final task is to expand the Jacobi elliptic function in powers
of $k_w$ when $k_w$ is small. A useful result is contained in
Ref.~\cite{AS}, and reads
\begin{equation}
\sn(u,k_w)\simeq\sin u+\frac{1}{3}\epsilon(\sin u\cos u-u)\cos u
\;,\label{4.2.16}
\end{equation}
in our case. The final step is to note that from
Eq.~(\ref{4.2.10}) we have
\begin{eqnarray}
u&=&(1+\xi)K(k_w)\nn\\
&\simeq&(1+\xi)\frac{\pi}{2}+\epsilon(1+\xi)\frac{\pi}{6}
+\cdots\;,\label{4.2.17}
\end{eqnarray}
so that $u$ in Eq.~(\ref{4.2.16}) also depends on $\epsilon$. If
we expand $\varphi(\xi)$ to order $\epsilon^{3/2}$ using the
expansions just described, after a short calculation we obtain a
result in exact agreement with Eq.~(\ref{4.2.9}).

The main conclusion of this section is that the approximation
method is in complete agreement with the expansion of the exact
result, at least for the first two orders in the expansion used.
This is sufficient to generate some faith in the general procedure
outlined in Sec.~\ref{sec3} in cases where it is not possible to
find an exact solution by analytical means. An example will be
given in the next subsection. It can also be used to provide a
useful check on the results of numerical calculations. We will
study such a case in Sec.~\ref{sec5}.

\subsection{Dirichlet boundary conditions in three dimensions}
\label{Dirichlet3D}

In this section we will examine the case of a cubical box of side
length $L$ with the field vanishing on all of the box walls.
Although this is perhaps a more realistic situation for a confined
field than that considered in Sec.~\ref{exact1D}, unfortunately we
have not been able to find the exact solution when the zero field
is unstable. This is quite unlike the situation where periodic or
antiperiodic boundary conditions are imposed on the walls. In this
case the exact solution in three dimensions can be simply related
to that found in one dimension~\cite{AvisIsham}. The Dirichlet
boundary conditions thwart the application of a similar procedure
here. However we can still use the approximation method described
in Sec.~\ref{sec3}.

The stability of the solution $\varphi=0$ is determined by the
lowest eigenvalue $\ell_0^2$ in Eq.~(\ref{1.10}). It is easy to
show that
\begin{equation}
\psi_0(x,y,z)=\cos\left(\frac{\pi x}{L}\right) \cos\left(\frac{\pi
y}{L}\right)\cos\left(\frac{\pi z}{L}\right) \label{4.3.1}
\end{equation}
is the eigenfunction of the Laplacian with the lowest eigenvalue
given by
\begin{equation}
\ell_0^2=\frac{3\pi^2}{L^2} \;.\label{4.3.2}
\end{equation}
The critical length $L_c$ is the value of $L$ for which
$\lambda_0$ defined in Eq.~(\ref{1.11}) vanishes. This gives
\begin{equation}
L_c^2=\frac{18\pi^2}{\lambda a^2}\;.\label{4.3.3}
\end{equation}
For $L<L_c$, $\varphi=0$ is stable, while for $L>L_c$ it is
unstable. Thus when $L>L_c$ the ground state will involve a
non-constant value of $\varphi$ in order to satisfy the Dirichlet
boundary conditions.

It proves convenient to define dimensionless coordinates as in
Eq.~(\ref{4.2.1})~:
\begin{equation}
x=\frac{L}{2}\,\xi_1\;,\ y=\frac{L}{2}\,\xi_2\;,\
z=\frac{L}{2}\,\xi_3\;, \label{4.3.4}
\end{equation}
so that the boundary of the cube is at $\xi_i=\pm1$ for $i=1,2,3$.
From Eqs.~(\ref{2.2}) and (\ref{2.3}) we find
\begin{eqnarray}
\nabla_c^2&=&\frac{4}{L_c^2}\,\sum_{i=1}^{3}
\frac{\partial^2}{\partial\xi_i^2}\;,\label{4.3.5a}\\
\nabla_1&=&-\frac{8}{L_c^2}\,\sum_{i=1}^{3}
\frac{\partial^2}{\partial\xi_i^2}\;,\label{4.3.5b}\\
\nabla_2&=&\frac{12}{L_c^2}\,\sum_{i=1}^{3}
\frac{\partial^2}{\partial\xi_i^2}\;.\label{4.3.5c}
\end{eqnarray}
A solution to Eq.~(\ref{2.5}) with the correct boundary conditions
is
\begin{equation}
\tilde{\varphi}_0(\xi_1,\xi_2,\xi_3)=\prod_{i=1}^{3}
\cos\left(\frac{\pi}{2}\xi_i\right)\;.\label{4.3.6}
\end{equation}
This may be used to calculate $C_1$ and $C_2$ in Eq.~(\ref{2.11})
and Eq.~(\ref{2.12}) resulting in
\begin{equation}
\frac{C_1}{2C_2}=\frac{128}{27}a^2\;.\label{4.3.7}
\end{equation}
The leading order approximation to the ground state therefore,
from Eq.~(\ref{2.14}), involves
\begin{equation}
\varphi_0(\xi_1,\xi_2,\xi_3)=\frac{8\sqrt{6}}{9}\,a
\prod_{i=1}^{3}
\cos\left(\frac{\pi}{2}\xi_i\right)\;.\label{4.3.8}
\end{equation}

To proceed to the next order we must solve the partial
differential equation Eq.~(\ref{2.6}) with Eq.~(\ref{4.3.8})
substituted for $\varphi_0$. A solution with the correct boundary
conditions can be shown to be
\begin{eqnarray}
\varphi_{1p}&=&-\frac{2\sqrt{6}}{81}\,a\Big\lbrace\frac{1}{3}
\cos\Big(\frac{3\pi}{2}\xi_1\Big)
\cos\Big(\frac{3\pi}{2}\xi_2\Big)
\cos\Big(\frac{3\pi}{2}\xi_3\Big)\nn\\
&&+\frac{3}{2}\Big\lbrack \cos\Big(\frac{\pi}{2}\xi_1\Big)
\cos\Big(\frac{3\pi}{2}\xi_2\Big)
\cos\Big(\frac{3\pi}{2}\xi_3\Big)\nn\\
&&\quad\quad +(1\leftrightarrow2)
+(1\leftrightarrow3)\Big\rbrack\nn\\
&&+9\Big\lbrack \cos\Big(\frac{\pi}{2}\xi_1\Big)
\cos\Big(\frac{\pi}{2}\xi_2\Big)
\cos\Big(\frac{3\pi}{2}\xi_3\Big)\nn\\
&&\quad\quad +(1\leftrightarrow3)
+(2\leftrightarrow3)\Big\rbrack\Big\rbrace \label{4.3.9}
\end{eqnarray}
where, to save space, we have used $(i\leftrightarrow j)$ to mean
the first term in the square brackets with the indices $i$ and $j$
on $\xi_i$ and $\xi_j$ switched.

Finally we evaluate $D_0$ and $D_1$ defined by Eq.~(\ref{2.19})
and Eq.~(\ref{2.20}). A straightforward calculation leads to
\begin{equation}
\frac{D_0}{D_1}=\frac{1375\sqrt{6}}{4374}\,a\;.\label{4.3.10}
\end{equation}
The solution may now be written down immediately from
Eq.~(\ref{2.22}) using Eqs.~(\ref{4.3.7}) and (\ref{4.3.10}) with
$\varphi_0$ given in Eq.~(\ref{4.3.8}) and $\varphi_{1p}$ in
Eq.~(\ref{4.3.9}). (To save space we will not write this out
explicitly.)

\section{The spherical cavity}
\label{sec5}

We now examine the case where the scalar field is confined by a
spherical shell of radius $R$, with the field vanishing at $r=R$.
The ground state should be spherically symmetric, so that
$\varphi$ will only depend on the radial coordinate $r$ if we use
the usual spherical polar coordinates. We need to solve
\begin{equation}
-\nabla_r^2\varphi+\frac{\lambda}{6}\varphi(\varphi^2-a^2)=0\;,\label{S1}
\end{equation}
where $\varphi=\varphi(r)$ and
\begin{equation}
\nabla_r^2=\frac{1}{r^2}\frac{d}{dr}\left(r^2\frac{d}{dr}\right)\;,
\label{S2}
\end{equation}
with $\varphi(r=R)=0$.

$\varphi=0$ is obviously a valid solution, and from the discussion
in Sec.~\ref{sec2}, the stability is determined once we know the
lowest eigenvalue of the Laplacian as in Eq.~(\ref{1.10}). In the
present situation it is easy to show that
\begin{equation}
\psi_0=\frac{\sin\left( \pi r/R\right)}{r}\;,\label{S3}
\end{equation}
is the eigenfunction of lowest eigenvalue, with
\begin{equation}
\ell_0^2=\left(\frac{\pi}{R}\right)^2\;.\label{S4}
\end{equation}
From Eq.~(\ref{1.11}), we can conclude that $\varphi=0$ is stable
for $R<R_c$ where the critical shell radius $R_c$ is defined by
\begin{equation}
R_c=\left(\frac{6}{\lambda}\right)^{1/2}\,\frac{\pi}{a}\;.
\label{S5}
\end{equation}
For $R>R_c$, $\varphi=0$ is not the ground state.

To find the stable ground state when $R>R_c$, we must solve the
non-linear differential equation Eq.~(\ref{S1}). We were unable to
find an exact analytical solution here, unlike the case of the
one-dimensional box where the differential equation was able to be
solved by quadrature. Instead we solved the equation by numerical
integration. The results are plotted in Fig.~3 for a range of
values of $R/R_c$.
\begin{figure}[htb]
\begin{center}
\leavevmode \epsfxsize=80mm \epsffile{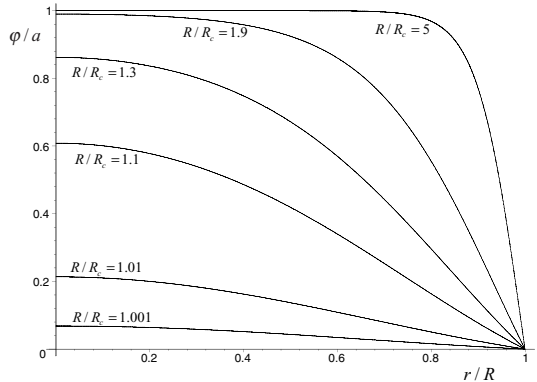}
\end{center}
\caption{The solution for $\varphi(r)/a$ is shown for various
values of $R/R_c$ as a function of $r/R$.}
\end{figure}
The results show that as the value of $R/R_c$ is increased, the
value of the field tries to get closer and closer to the constant
value of $\varphi=a$ over a greater range of values of the radius.
Because the boundary conditions require $\varphi(r=R)=0$, this
results in a very sharp drop-off in the value of the field as
$r\rightarrow R$. The profile of the field starts to approach a
step-function as $R/R_c\rightarrow\infty$.

Although we were not able to find an analytic solution to
Eq.~(\ref{S1}), we can still apply the approximation method
described in Sec.~\ref{sec3}. A dimensionless radial coordinate
$\rho$ is defined by
\begin{equation}
r=R\rho\;, \label{S6}
\end{equation}
and we set
\begin{equation}
R=(1+\epsilon)R_c\;\label{S7}
\end{equation}
as in Eq.~(\ref{2.2}). With $\nabla^2$ given by Eq.~(\ref{S2}) we
find the expansion Eq.~(\ref{2.3}) where
\begin{eqnarray}
\nabla_c^2&=&\frac{1}{R_c^2}\nabla_{\rho}^{2}\;,\label{S8a}\\
\nabla_1&=&-\frac{2}{R_c^2}\nabla_{\rho}^{2}\;,\label{S8b}\\
\nabla_2&=&\frac{3}{R_c^2}\nabla_{\rho}^{2}\;,\label{S8c}
\end{eqnarray}
and $\nabla_{\rho}^{2}$ given by taking $r=\rho$ in
Eq.~(\ref{S2}).

The lowest order contribution to the approximate solution follows
from solving Eq.~(\ref{2.5}). This results in
\begin{equation}
\tilde{\varphi}_0(\rho)=\frac{\sin(\pi\rho)}{\rho} \;. \label{S9}
\end{equation}
The overall scale is set by calculating $C_1$ and $C_2$ in
Eqs.~(\ref{2.11}) and (\ref{2.12}) with the result that
\begin{equation}
\left(\frac{C_1}{2C_2}\right)^{1/2}=\frac{a}{\sqrt{I}}
\;,\label{S10}
\end{equation}
with~\cite{SI}
\begin{equation}
I=\frac{\pi}{2}\lbrack \Si(2\pi)-\Si(4\pi)\rbrack \;.\label{S11}
\end{equation}
This gives
\begin{equation}
\varphi_0(\rho)=\frac{a\sin(\pi\rho)}{\sqrt{I}\,\rho}
\;,\label{S12}
\end{equation}
to be used in Eq.~(\ref{2.6}).

A particular solution to Eq.~(\ref{3.6}) can be shown to
be~\cite{SI}
\begin{eqnarray}
\varphi_{1p}(\rho)&=&-\frac{\pi^2 a}{2I^{3/2}}\lbrack
\Ci(4\pi\rho)-\Ci(2\pi\rho)\rbrack\frac{\sin(\pi\rho)}{\rho}\nn\\
&&+ \frac{\pi^2 a}{2I^{3/2}}\lbrack
\Si(4\pi\rho)-2\Si(2\pi\rho)\rbrack\frac{\cos(\pi\rho)}{\rho}\nn\\
&&+\frac{\pi a}{I^{1/2}}\cos(\pi\rho)\;.\label{S13}
\end{eqnarray}
The complicated nature of this renders the calculation of $D_0$ in
Eq.~(\ref{2.19}) difficult, although $D_1$ in Eq.~(\ref{2.20}) is
easily evaluated. It can be shown that
\begin{equation}
\frac{D_0}{D_1}\simeq0.417\,349\,770\;a \;.\label{S14}
\end{equation}
This is sufficient to determine the approximate solution in
Eq.~(\ref{2.22}) correct to order $\epsilon^{3/2}$.

As a check on the numerical results shown in Fig.~3, we plotted
the approximate solution we have just described. For small
$\epsilon$ the result was found to be indistinguishable from the
result of numerical integration of Eq.~(\ref{S1}); however, as
$\epsilon$ is increased the agreement becomes less good as would
be expected for a small $\epsilon$ expansion. The disagreement is
largest near $\rho=0$ where the field has its largest value. To
get an idea of how close the approximate solution is to the true
result, some of the values found for the field at the origin are
included in Table~1.
\begin{table}[t]
\begin{ruledtabular}
\begin{tabular}{|c|c|c|c|}
 $\epsilon$&$\varphi(0)/a$&order $\epsilon^{1/2}$&
 order $\epsilon^{3/2}$\\
\hline 0.001&0.068291&0.068370&0.068288\\
0.01&0.213584&0.216206&0.213590\\
0.1&0.608397&0.683703&0.601004\\
0.2&0.774283&0.966902&0.732994\\
0.3&0.862136&1.184208&0.754491
\end{tabular}
\end{ruledtabular}
\caption{A table showing the value of the field at the origin in
units of $a$. $\epsilon$ shows how close $R$ is to $R_c$ as in
Eq.~(\ref{S7}). The second column gives the true value of
$\varphi(0)/a$ found by numerical integration of Eq.~(\ref{S1}).
The final two columns show respectively the results found from the
approximate solution Eq.~(\ref{2.22}) using the leading order
term, and the next order correction.}
\end{table}
It is clear that by including the correction to the leading term a
more accurate result is obtained for a wider range of $\epsilon$.
Even for relatively large values of $\epsilon$ ($\epsilon\simeq1$)
where the small $\epsilon$ expansion would not be expected to be
particularly accurate, we found the approximate solution to be
close to the true one for $\rho$ close to $\rho=1$, corresponding
to the radius of the shell. Thus the approximation method
described in Sec.~\ref{sec3} can provide a useful check on the
results of numerical integration in cases where no exact solution
is known.

\section{Summary and conclusions}
\label{sec6}

We have presented an analysis of the ground state for a real
scalar field confined in a cavity.  Although we concentrated on
the double-well potential (2.2) it would be easy to extend the
analysis to other potentials.  (For example, in the case the
bi-cubic potential, an exact solution in the cavity can be found
in terms of the Weierstrass elliptic function as in
Ref.~\cite{HuishToms}.) In addition, the analysis can be extended
to curved space, although it would be difficult to find exact
solutions except in special cases.

We presented a method for obtaining approximate analytical
solutions in the case where the length scales of the cavity were
close to the critical values at which $\varphi=0$ became unstable.
For the rectangular cavity, we showed that this approximation
method agreed with an expansion of the exact solution which we
found. When $\varphi=0$ was not the ground state, it was shown
that as the ratio of the size of the cavity to the critical size
$L/L_c$ increases, the ground state tends towards the constant
value which minimises the potential over as much of the cavity as
possible. The condition that the field vanishes on the box walls
means that the field must always drop to zero, with the drop-off
becoming increasingly sharp as $L/L_c$ is increased.  A similar
behaviour was found for the case of a spherical cavity.  By
extrapolation, for a field confined to vanish on the walls of a
general cavity, if the cavity is sufficiently large it would be
expected that the ground state would correspond to a field which
was constant almost everywhere inside the cavity, with a very
sharp drop-off to zero as the cavity boundary is approached.

Although we have concentrated on fields confined by cavities in
the present paper, the general method described above can be
useful in other situations where the boundary conditions prohibit
constant values of the field.  We plan to report on this
elsewhere.


\begin{thebibliography}{99}
\bibitem{SI} Here $\Si(x)$ denotes the Sine integral,
defined by $\displaystyle{\Si(x)= \int_{0}^{x}dt\,\frac{\sin
t}{t}}$. For later use we also note the Cosine integral defined by
$\displaystyle{\Ci(x)=\gamma+\ln\,x+ \int_{0}^{x}dt\,\frac{(\cos
t-1)}{t}}$.
\end{thebibliography}
\end{document}